\documentclass[iop,numberedappendix,appendixfloats]{emulateapj}
\shortauthors{Yan, Ma, Beacom, \& Runge}

\usepackage{natbib}
\usepackage{enumitem}
\usepackage{amsmath}
\usepackage{soul}

\usepackage[bookmarks=true, pagebackref=true]{hyperref}

\newcommand{\Herschel}{\textit{Herschel}}
\newcommand{\HST}{\textit{HST}}
\newcommand{\Spitzer}{\textit{Spitzer}}
\newcommand{\Chandra}{\textit{Chandra}}
\newcommand{\JWST}{\textit{JWST}}
\newcommand{\up}{${u}^{\prime}$}
\newcommand{\gp}{${g}^{\prime}$}
\newcommand{\rp}{${r}^{\prime}$}
\newcommand{\ip}{${i}^{\prime}$}
\newcommand{\zp}{${z}^{\prime}$}

\begin{document}

\title{Revealing Dusty Supernovae in High-Redshift (Ultra-)Luminous InfraRed
Galaxies Through Near-Infrared Integrated Light Variability}

\author{Haojing Yan \altaffilmark{1}}
\email{yanha@missouri.edu}

\author{Zhiyuan Ma \altaffilmark{1}}
\email{zmzff@mail.missouri.edu}

\author{John F. Beacom \altaffilmark{2,3,4}}
\email{beacom.7@osu.edu}

\author{James Runge \altaffilmark{1}}
\email{jmr24f@mail.missouri.edu}

\altaffiltext{1}{Department of Physics and Astronomy, University of Missouri-Columbia, USA}
\altaffiltext{2}{Department of Physics, The Ohio State University, USA}
\altaffiltext{3}{Department of Astronomy, The Ohio State University, USA}
\altaffiltext{4}{Center for Cosmology and AstroParticle Physics (CCAPP), The Ohio State University, USA}

\begin{abstract}

   Luminous and ultra-luminous infrared galaxies ((U)LIRGs) are rare today but
are increasingly abundant at high redshifts. They are believed to be dusty
starbursts, and hence should have high rates of supernovae (multiple events per
year). Due to their extremely dusty environment, however, such supernovae could
only be detected in rest-frame infrared and longer wavelengths, where our
current facilities lack the capability of finding them individually beyond the
local universe. We propose a new technique for higher redshifts, which is to
search for the presence of supernovae through the variability of the integrated
rest-frame infrared light of the entire hosts. We present a pilot study to
assess the feasibility of this technique. We exploit a unique region, the
``IRAC Dark Field'' (IDF), that the {\it Spitzer Space Telescope} has observed
for more than 14 years in 3--5~\micron. The IDF also has deep far-infrared data 
(200--550~\micron) from the {\it Herschel Space Observatory} that allow us to
select high-redshift (U)LIRGs.  We obtain a sample of (U)LIRGs that have secure 
optical counterparts, and examine their light curves in 3--5~\micron. While the
variabilities could also be caused by AGNs, we show that such contaminations
can be identified. We present two cases where the distinct features in their
light curves are consistent with multiple supernovae overlapping in time. 
Searching for supernovae this way will be relevant to the James Webb Space 
Telescope (\JWST) to probe high-redshift (U)LIRGs into their nuclear regions
where \JWST\, will be limited by its resolution.

\end{abstract}

\keywords{galaxies: active; galaxies: starburst; infrared radiation; (ISM:) dust, extinction; (stars:) supernovae: general}

\section{Introduction}

 
  Luminous and ultra-luminous infrared galaxies (LIRGs and ULIRGs; hereafter
``(U)LIRGs'') have high IR luminosity (integrated over rest-frame 
8--1000~\micron) of $L_{IR}>10^{11}$ and $>10^{12}L_\odot$, respectively. It is
widely believed that (U)LIRGs are starbursts, and that their strong IR
emissions are mostly due to the dust-reprocessed UV photons from their large
numbers of young stars. On the other hand, (U)LIRGs usually also harbor AGNs,
which often makes it difficult to attribute their major IR power source solely
to starbursts \citep[see][]{Lonsdale2006}. It is important to distinguish these
hypotheses or, more likely, to determine their fractional contributions to the
(U)LIRG power and how those may change with redshift. Although (U)LIRGs are 
rare today, they are more common at high redshifts
\citep[e.g.,][]{LeFloch2005, Magnelli2013}, which make them relevant to the
global picture of galaxy evolution across cosmic time.

  A decisive way to determine the fraction of (U)LIRGs' power supplied by 
starbursts would be to measure their rates of supernovae (SNe). If they are 
indeed dominantly powered by starbursts, (U)LIRGs should have high star 
formation rates (SFRs; $>10$ and $>100M_\odot$~yr$^{-1}$ for LIRGs and ULIRGs, 
respectively), and hence should also have high rates of SNe. For ULIRGs, the 
rate $r_{SN}$ is estimated to be $\gtrsim 2$--3 events yr$^{-1}$
\citep[][]{vanBuren1994, Mattila2001}. However, due to the severe dust 
obscuration, such SNe would have to be searched for in the radio or IR. 

   VLBI detections of radio SNe and SN remnants in the ULIRG 
Arp 220 and the LIRG Arp 299 \citep[e.g.,][and references therein]
{Lonsdale2006b, PerezTorres2009, RomeroCanizales2011, RomeroCanizales2014, Bondi2012, Varenius2017}
lend strong support to high $r_{SN}$ in (U)LIRGs. For example, 
\citet[][]{Lonsdale2006b} derive $r_{SN} \sim 4$ events yr$^{-1}$ for Arp 220,
which implies a sufficient SFR to account for its $L_{IR}$ without resorting
to AGN. VLBI imaging, however, still cannot be applied to (U)LIRGs beyond the
local universe.
IR surveys have also discovered a few tens of SNe in local (U)LIRGs 
\cite[e.g.,][]{Maiolino2002, Cresci2007, Mattila2007, Kankare2008, Miluzio2013, 
Kool2018}, including multiple SNe in the same galaxies
\citep[e.g.,][]{Kankare2012, Kankare2014}. Two main conclusions have
been drawn from these results: (1) (U)LIRGs indeed have a high rate of
SNe embedded by dust, which are not visible to optical surveys. (2) The current
IR surveys must still be missing a large fraction of dusty SNe close to the
nuclear region of the host galaxies due to both the extreme dust extinctions
and the much decreased survey sensitivities when working against the bright
background. Unfortunately, these IR surveys also do not go beyond the local
universe.

  As an alternative to discovering SNe individually, we propose to reveal them
through the variability of the integrated IR light of the host. This method can
be easily applied to high-$z$, because it requires only high-precision 
differential photometry. As far as we are aware, no search of this type
has been performed, and a pilot study is needed to assess the prospects for
present instruments and especially for the {\it James Webb Space Telescope}
(\JWST). To set the scale required, a supernova peaking at
$M\sim -19$~mag within a host of $M\sim -21$~mag (i.e., $m\sim 22.4$~mag at
$z\approx 1$) would increase the host brightness by $\sim 0.16$~mag, which
should be detectable by current facilities. If a (U)LIRG has multiple SNe per
year, they could overlap in time and result in even larger variability. 
Admittedly, finding evidence of dust-embedded SNe in this way is inferior
to resolving them individually in terms of the follow-up applications of the
SNe; nevertheless, it is still a powerful means to probe the nuclear regions
that the activities are expected to be the most violent and yet are the most
difficult to penetrate. The only major contaminations would be AGNs, which are
known to vary with typical amplitudes of a few tenths of magnitude
\citep[e.g.,][]{Peterson2001}. 

   In this paper,  we exploit a unique field known as the ``IRAC Dark Field''
\citep[hereafter ``IDF''; see][]{Krick2009}. Since its launch in 2003, the
{\it Spitzer Space Telescope} has been observing this area for the calibration
of its InfraRed Array Camera (IRAC; \citet[][]{Fazio2004}), producing a deep
field of $\sim 13$\arcmin\, in radius. The IDF (R.A. $=17^{\rm h}40^{\rm m}$, 
decl. $=68^{\rm o}40^{\prime}$, J2000) is close to the North Ecliptic Pole, and 
is in a region of the lowest zodiacal background (hence ``dark''). During the
cryogenic phase (2003 October to 2009 May), all four IRAC channels (3.6, 4.5,
5.8, and 8.0~\micron) were used. After the coolant depletion (``warm-mission''
phase), the 3.6 and 4.5~\micron\, channels (hereafter ``Ch1'' and ``Ch2,'' or
``Ch1/2'') have continued observations without any significant loss of
sensitivity. The IDF is the only region on the sky with long-duration 
($>14$ years) monitoring data in 3--5~\micron\, (sampling rest-frame near-IR up
to $z\approx 3.5$), and its early data ($\sim 2$ years) were already of
unprecedented time baseline such that it inspired a search for Population III
supernovae, albeit with null results \citep[][]{Frost2009}. The IDF is ideal
for our purpose here because of two additional reasons: it has (1) hundreds of 
(U)LIRGs revealed by the far-IR (FIR) observations from the {\it Herschel Space
Observatory} and (2) medium-deep \Chandra\, X-ray observations for AGN
diagnostics. 

   Our paper is organized as follows. We present the data in \S 2, and describe
the selection of variable objects in \S 3. The analysis of these variable
objects is given in \S 4, which is followed by a discussion in \S 5. A brief
summary is given in \S 6. We use AB magnitudes throughout the paper, and adopt
$\Omega_M=0.27$, $\Omega_\Lambda=0.73$ and $H_0=71$~km s$^{-1}$ Mpc$^{-1}$.

\section{Data and Analysis}

  We describe the data used in this study, which span a wide range from X-ray
to FIR.

\subsection{Spitzer IRAC Data}

   Our analysis is based on the IRAC Ch1/2 images from 2003 October through
2017 December. The retrieved data include the ``Basic Calibrated Data'' (BCDs),
which are single exposures with the major instrumental effects removed by the
standard \Spitzer\, Science Center (SSC) data reduction pipeline, and the 
so-called ``post-BCD'' (PBCD) products,
which are the combined results from the BCDs within a single observation 
sequence known as a ``Astronomical Observation Request'' (AOR). 
As the observations in Ch1 and Ch2 are simultaneous (but
in two adjacent fields), the PBCD products of each AOR contain the mosaics
(and other diagnostic files) in both Ch1 and Ch2.

   The AOR designs depend on the goals of the calibrations, and thus are not
uniform: they can be different in the frame time of single exposures, the total
duration, the field position, the spatial coverage, etc. We only kept the AORs
whose single exposures have a frame time $\geq 100$ seconds. This resulted in
424 and 635 AORs in the cryogenic and the warm-mission phases, respectively,
i.e., a total of 424+635 $=$ 1059 PBCD mosaics at 1059 epochs in both Ch1 and
Ch2. Figure 1 shows the epoch map of these AORs. These mosaics have median
integration times ranging from 96.8 to 2516.8 s/pixel, and the majority have
either 374.4--387.2 s/pixel (30.8\%) or 561.6--580.8 s/pixel (46.4\%) in Ch1,
and either 387.2 s/pixel (30.8\%) or 484.0 s/pixel (41.1\%) in Ch2,
respectively. The pixel scale of these PBCD products is 0\arcsec.6,
which is about half of the native pixel sizes.

\begin{figure}[]
\plotone{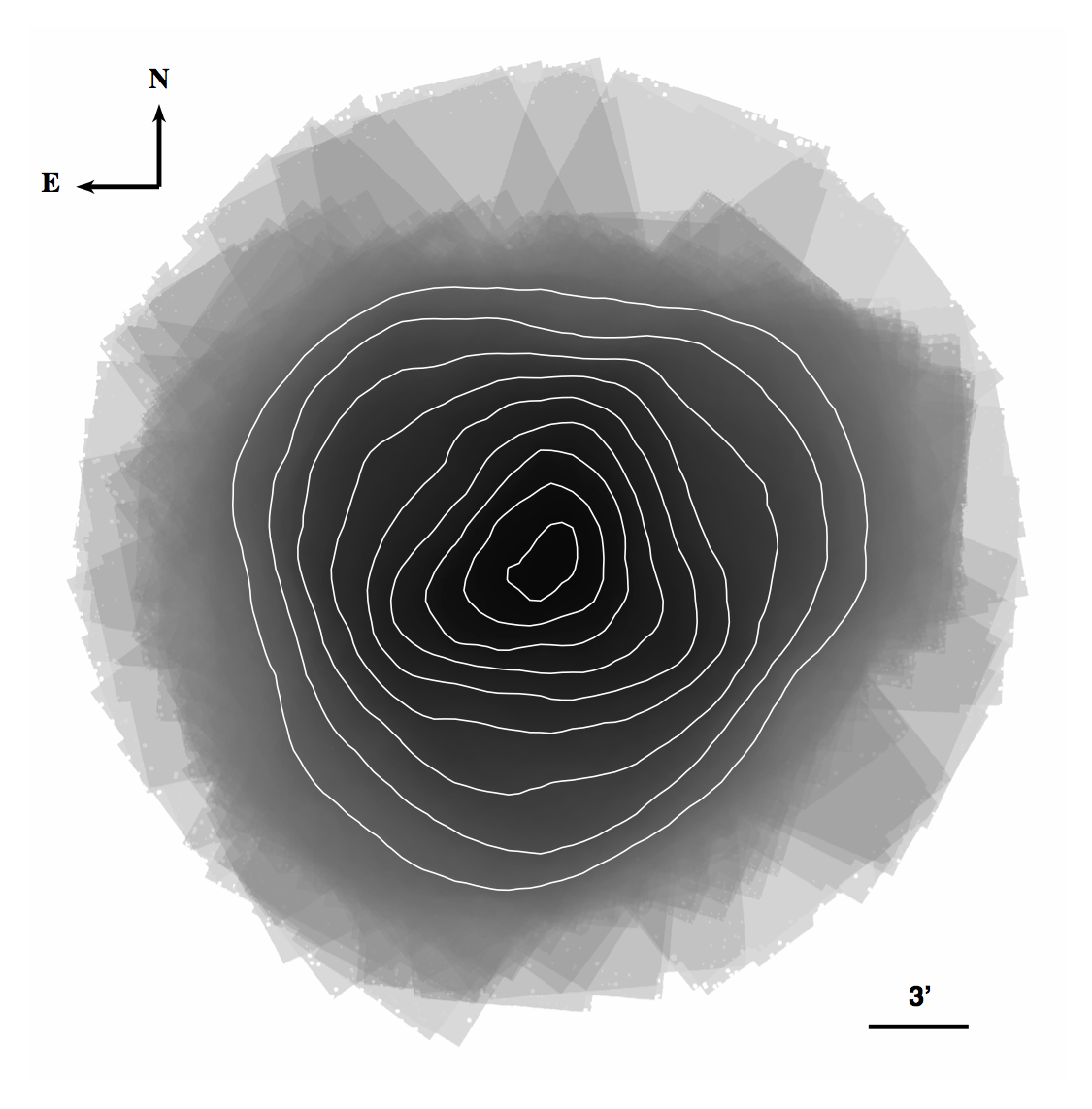}
\caption{IRAC Ch1 epoch map in the IDF ($\sim 13$\arcmin\, radius), based on
the 1059 AORs (\S 2.1). The contours are shown from 100 to 900
epochs in a 100 epoch step-size. The Ch2 epoch map is similar.
}
\label{fig:ch1epochmap}
\end{figure}

   We ran SExtractor \citep[][]{Bertin1996} on each PBCD mosaic for
photometry, using the associated uncertainty map (i.e., specifying the flux
uncertainty at each pixel) as the RMS map. We adopted \texttt{MAG\_{AP}} with 
circular apertures of 2\arcsec, 3\arcsec, 4\arcsec, 5\arcsec, and 6\arcsec\,
in diameter. The individual source catalogs were matched to generate the 
``light-curve catalog'' of 1059 epochs in Ch1/2. We found that the data in
the first 37 AORs of the warm-mission phase (MJD 55035 through 55115) likely
suffered from photometric zero-point errors, and hence excluded these data.
The final number of useful epochs is 1022.

\subsection{Herschel SPIRE Data}

  The \Herschel\, Spectral and Photometric Imaging Receiver 
\citep[SPIRE;][]{Griffin2010} periodically observed the IDF region 
(in the 250, 350, and 500~\micron\, bands) for calibration throughout
its mission (from 2009 September to 2013 April). In total, there are 151 
AORs of various coverage and central pointings, mostly being large or small
mapping scans with durations ranging from about eight minutes to two hours.

   We retrieved and analyzed these data using the \Herschel\, Interactive 
Processing Environment \citep[HIPE;][]{Ott2010}. The instrumental effects have
already been removed by the \Herschel\, data reduction pipeline, which result
in the ``Level 1'' products. Our reduction was to stack these products of the
151 AORs. SPIRE always observed its three bands in the same field of view, and
hence we obtained the mosaics and their noise maps in all three bands. The 
final mosaics extend an area of $\sim 0.88$~deg$^2$, with the pixel scales of
6\arcsec, 10\arcsec, and 14\arcsec\, in 250, 350, and 500~\micron, respectively.

  We generated a band-merged catalog using 250~\micron\, as the detection band,
following the general procedure of \citet[]{Wang2014}. The detection was done
on the 250~\micron\, map using StarFinder \citep[SF;][]{Diolaiti2000}, which is
an iterative source finding program that can deal with images of significant
source blending problem. The detection was done iteratively using a PSF-fitting 
technique so that faint sources around bright ones could be included.
We adopted the following SF parameters: ``\texttt{SNR\_thresh}'' of 1.5,
``\texttt{Correction thresh}'' of 0.7, and ``\texttt{Deblending distance}'' of
0.7$\times$FWHM, where FWHM was set to 18\arcsec.15 for the 250~\micron\,
band.  The positions of these 250~\micron\, sources were sent to the task
\texttt{sourceExtractorSimultaneous} in HIPE to do photometry in all three 
bands simultaneously. We used a Gaussian Point Response Function (PRF) with the
FWHM value set to 18\arcsec.15, 25\arcsec.15, and 36\arcsec.30 at 250, 350, and
500~\micron, respectively \citep[see][]{Swinyard2010}. The routine generated
flux densities, as well as their uncertainties based on the noise maps. For the
latter, we added in quadrature a constant confusion noise of 5~mJy to obtain
the final estimates of the uncertainties. In total, we detected 1759 SPIRE
sources within the IRAC coverage, 208 of which have S/N $\geq 5$.

\subsection{Chandra X-ray Data}

   The IDF has been observed by the \Chandra\, ACIS-I camera for $\sim 100$~ks,
and these observations are described in \citet[][]{Krick2009}. Following the
procedures similar to theirs, we reduced these data independently using the 
Chandra Interactive Analysis of Observations software (CIAO, v4.9; with CALDB
4.7.0). All observations were reprocessed using the \texttt{chandra\_repro}
script, which corrects for image defects (such as hot pixels and cosmic-ray 
afterglows) and does background cleaning before creating a final event list.
The processed event files for each observation were then merged using
\texttt{merge\_obs}, similar to \texttt{merge\_all} used by 
\citet[][]{Krick2009}
except that it folds in the \texttt{reproject\_aspect} script. We then ran
\texttt{wavdetect} on the merged event list with the ``mexican hat'' wavelet
functions on size scales from 1 to 8. The merged event list was converted
into flux images through the use of \texttt{eff2evt} in three energy bands
(soft: 0.5--1.2~keV; medium: 1.2--2.0~keV; hard: 2.0--7.0~keV). The positions
from the \texttt{wavdetect} run were fed to \texttt{roi} to create a source and
background region for each object. Source and background measurements were then
made on the individual flux images using \texttt{dmstat}. In total, we
extracted 121 sources within the IRAC coverage. 

\subsection{WIYN and HST Optical Data}

   We have been observing the IDF using the One-Degree Imager (ODI) in 
\up\gp\rp\ip\zp\, at the WIYN telescope. For this study,
we used the \ip-band data obtained on 2017 March 29 and April 1,
with the purpose of providing the positional priors
for the SPIRE source counterpart identification. These images, totaling
two hours of integration, were first reduced by the ODI pipeline to remove the
instrumental effects and to calibrate their astrometry based on the GAIA 
Data Release 1. We then stacked them using the SWarp software (by E. Bertin;
V2.38.0). The mosaic has reached 25.2 mag (5 $\sigma$), has the PSF FWHM of
0\arcsec.65, and has an rms accuracy of 30 mas in astrometry.

   The IDF has also been observed by the \HST\, Advanced Camera for Surveys
(ACS) in the F814W-band at 2-orbit depth \citep[see][]{Krick2009}. Here, we only
use them to obtain the morphologies of interesting sources when necessary, and
hence we rely on the per-visit stacks (pixel scale 0\arcsec.05) contained in
the pipeline-reduced data directly retrieved from the \HST.

\section{Searching for Variability}

   As our interest is the SNe of the (U)LIRGs and the possible contamination
due to AGNs, here we limit our variability search to only the IRAC counterparts
of the SPIRE sources (potential (U)LIRGs) and the \Chandra\, sources
(potential AGNs). 

\subsection{Optical/IR Counterparts of SPIRE Sources and Chandra Sources}

  We first identified the optical counterparts of the aforementioned 208 
S/N $\geq 5$ SPIRE 250~\micron\, sources, using our own Counterpart Identifier
tool (``CIDer''; Z. Ma \& H. Yan, in preparation) developed following the
general PSF-fitting methodology as in \citet[][]{Yan2014}. For a given 
250~\micron\, source, which might be the blended product of multiple objects,
the ODI \ip-band image is used to locate the possible contributors to its flux.
CIDer only identifies the major contributors, i.e., the \ip-band objects
that contribute the bulk of the 250~\micron\, flux. This is
achieved by iteratively fitting the 250~\micron\, PSF at the positions of the
potential contributors, which is appropriate because the 250~\micron\, image has
so coarse a resolution (PSF FWHM $\sim 18$\arcsec) that all sources are
point-like. We required that
a major contributor must contribute $\geq 25$\% of the 250~\micron\, flux. 
In total, CIDer identified
201 \ip-band objects as the major contributors to 148 250~\micron\, sources.
We then cross-matched these \ip-band objects with those in the IRAC light-curve
catalog, using a matching radius of 1\arcsec. All these 201
\ip-band objects have IRAC counterparts.

   The identification of the IRAC counterparts of the 121 \Chandra\, sources was
straightforward, which was done by cross-matching to the IRAC light-curve
catalog using a matching radius of 1\arcsec. In total, 71 \Chandra\, sources
have IRAC counterparts.

\begin{figure*}[t]
\plotone{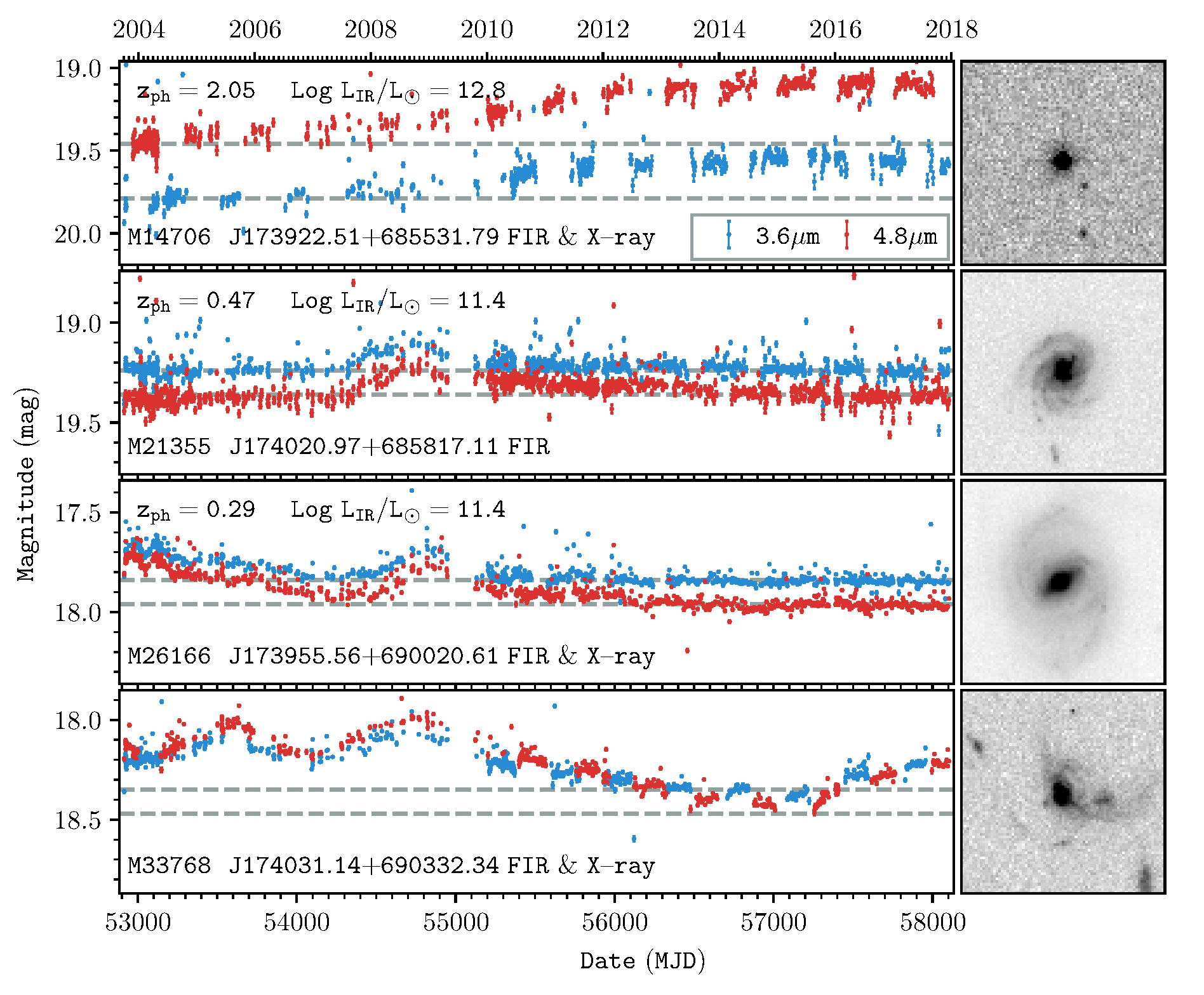}
\caption{IRAC Ch1 (blue) and Ch2 (red) light curves of the four variable 
sources in the FIR sample (indicated by ``FIR''), with their IDs and names
(IAU convention) noted. The three objects that are also in the X-ray sample
are labeled by ``X-ray.'' The magnitudes are based on circular apertures of 
4\arcsec\, diameter, and aperture corrections of $-0.36$ and $-0.39$~mag have
been applied to Ch1 and Ch2, respectively. The \HST\, F184W image stamps
(6\arcsec$\times$6\arcsec) to the right show their morphologies. 
The first three objects have $z_{ph}$ and therefore their $L_{IR}$ can be
calculated (see \S 3.4), Figure 4 and Figure 5), and these values are also labeled.
The last one currently does not have $z_{ph}$.
}
\label{fig:FIRlc}
\end{figure*}

\begin{figure*}[t]
\plotone{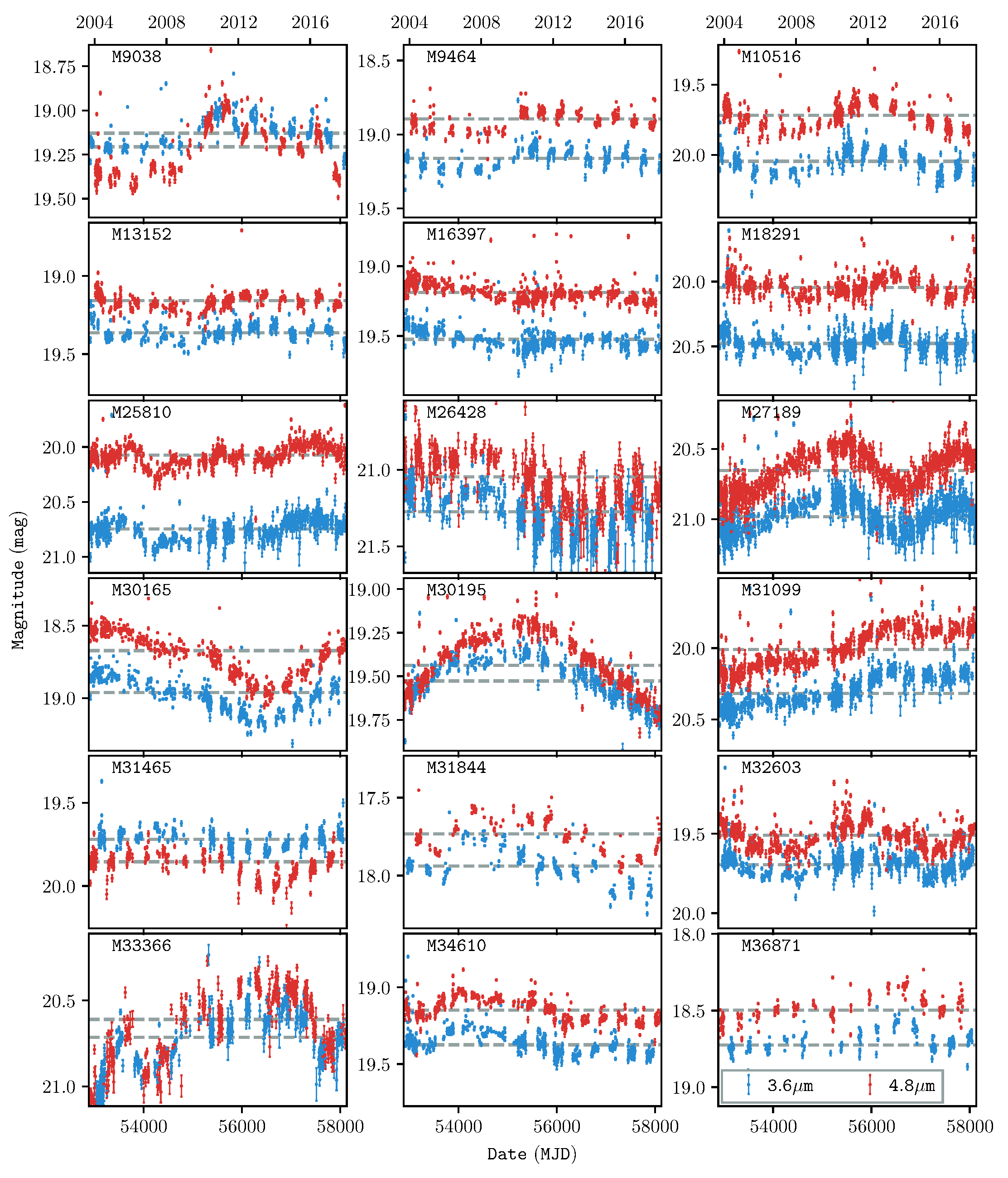}
\caption{IRAC Ch1 (blue) and Ch2 (red) light curves of the 18 objects in the
X-ray sample (with their ID noted), excluding the three already shown in
Figure 2. Magnitudes are the same as in Figure 2.
}
\label{fig:Xraylc}
\end{figure*}

\subsection{FIR Sample, X-ray Sample, and Control Sample}

  To ensure the most reliable detection of IRAC variability,
we require that an object must have photometry over at least 100
epochs in either Ch1 alone or Ch2 alone. The variability was searched through
these three samples:

  (1) Among the 201 \ip-band major contributors to 148 
\Herschel/SPIRE 250\micron\, sources (i.e., potential (U)LIRGs) within the IRAC
coverage, 96 have $\geq 100$ epochs of photometry and thus comprise our
FIR sample.

  (2) Among the 71 X-ray sources (i.e., potential AGNs) within the IRAC
coverage that have IRAC counterparts, 67 satisfy the same criterion and
hence comprise our X-ray sample. There are 8 objects in common with the FIR
sample.

  (3) A control, ``field'' sample is constructed by selecting the IRAC sources
that are $>21$\arcsec\, away from any SPIRE 250~\micron\, sources
and $>3$\arcsec\, away from of any X-ray sources. This includes 1987
objects that have $\geq 100$ epochs of photometry.

\subsection{Variable Objects}

   We adopted the $d=4$\arcsec\, \texttt{MAG\_AP} values as the benchmark, and
used those measured in other apertures for verification once a candidate was
found. A candidate variable source was selected using these criteria: (1) its
brightness continuously changes over $>30$ days, (2) the peak-to-valley 
variation is $> 0.1$~mag, and (3) the average photometric error over the
changing period is $<0.05$~mag.  In addition, its light curve and images were
visually examined to confirm its legitimacy. 

   In the end, 4 and 21 variable objects were found in the FIR and the 
X-ray samples, respectively. Three of these objects are in common, resulting 22
unique sources in total. In contrast, none of the 1987 objects in the control
sample have {\it comparable} variability. Figure 2 shows the light curves and the
ACS F814W images of the variables in the FIR sample. For comparison, Figure 3
shows the light curves of the 18 variables in the X-ray sample (i.e., 
excluding the three duplicates from the FIR sample). Note that these Ch1/2
magnitudes are based on $d=4$\arcsec\, aperture but have been applied the
corrections of $-0.102$ and $-0.105$~mag in Ch1 and Ch2, respectively, to
convert to the ``total'' magnitudes. These values are obtained by interpolating
between the aperture corrections of $d=3\arcsec.6$ and $d=4\arcsec.8$ provided
in the IRAC Instrument Handbook.



\begin{figure*}[t]
\plotone{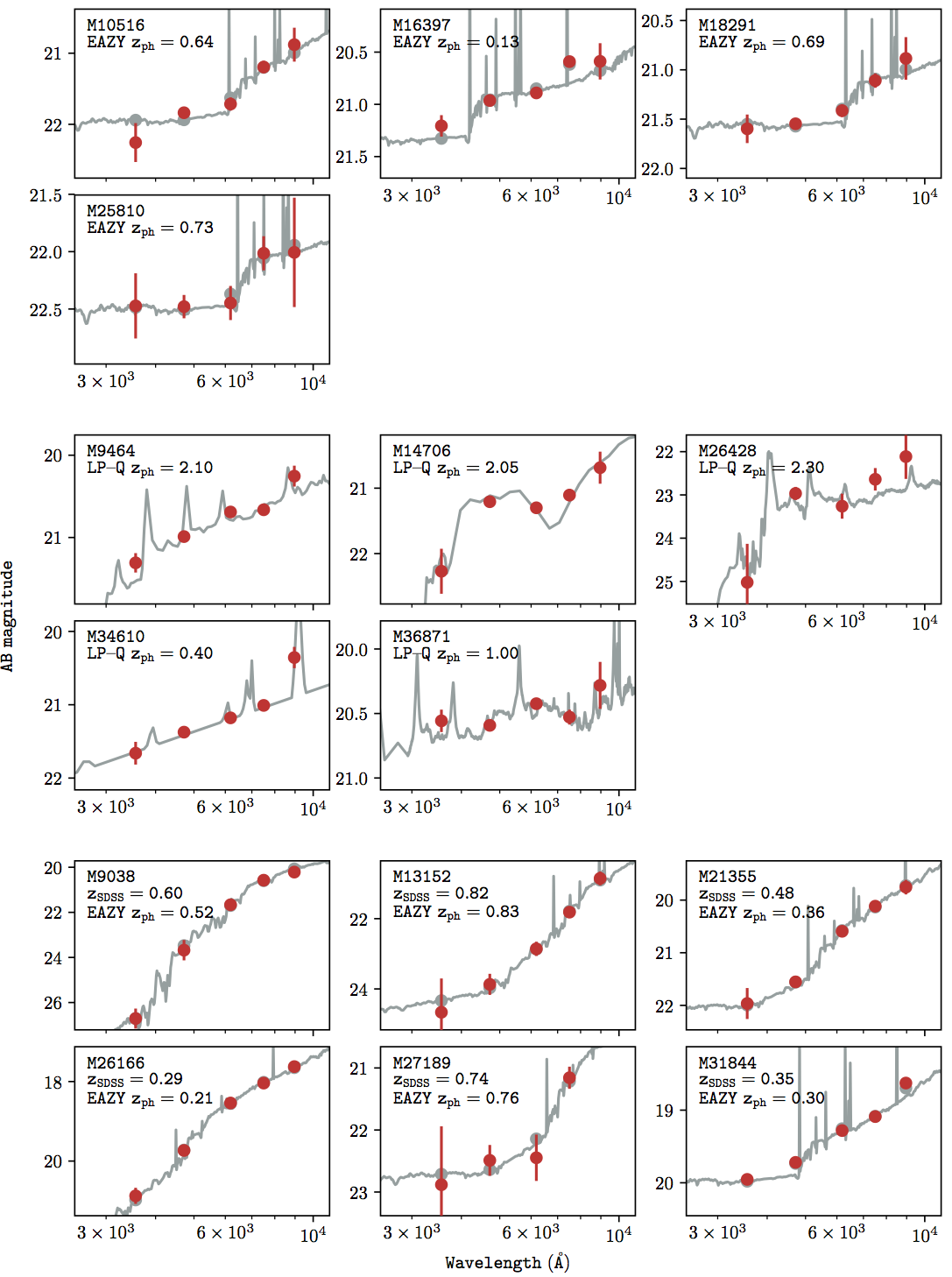}
\caption{Optical SED fitting and $z_{ph}$ derivation for the 15 objects that
have SDSS photometry. The red dots with error bars are the data points,
and the gray curves are the best-fit models. The
first four rows show the nine objects that do not have existing SDSS $z_{ph}$,
for which we derive $z_{ph}$ independently. Among these, four objects (shown in
the first two rows) can be fit by galaxy models using EAZY (results labeled as
``EAZY $z_{ph}$''), and the other five (shown in the next two rows) can only be
fit by AGN/QSO models using LePhare (results labeled as ``LP-Q $z_{ph}$''). To
check the consistency, the six objects with existing SDSS $z_{ph}$ (shown in
the last two rows and values labeled with ``$z_{SDSS}$'') are also fit using
the same procedure, and they can all be fit by galaxy models using EAZY. The
mean difference between the two sets is $<$$\Delta z/(1+z)$$>$$=0.04$.
}
\label{fig:fig_zph}
\end{figure*}

\begin{figure*}[t]
\plotone{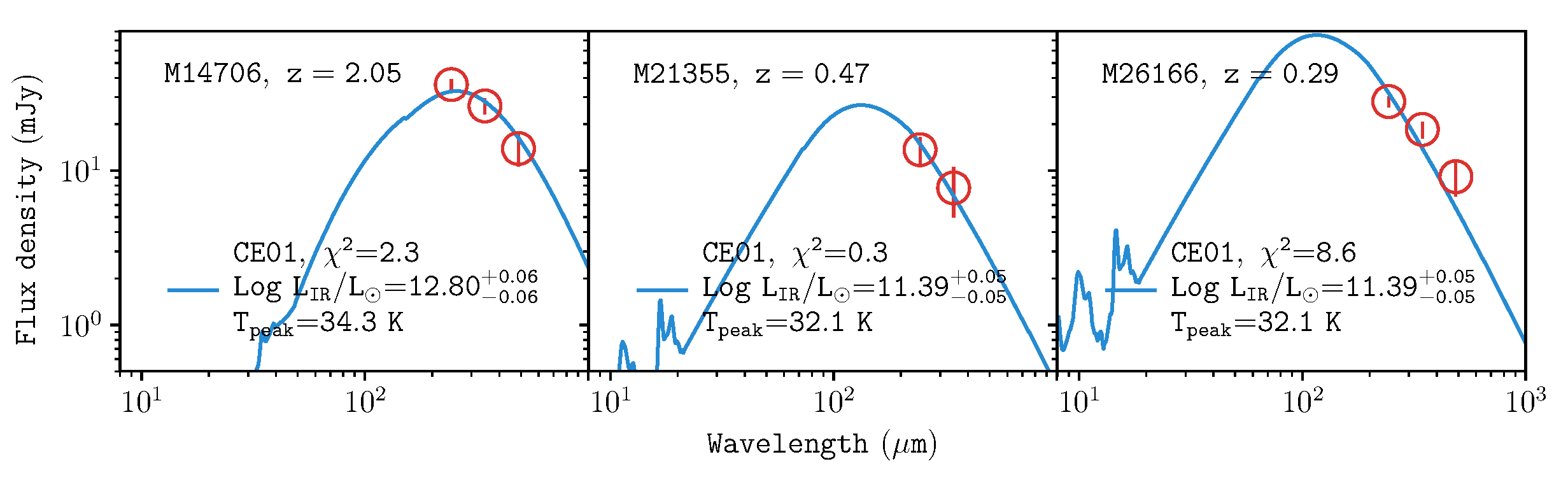}
\caption{FIR SED fitting of the three objects in the FIR sample that have
$z_{ph}$. The red circles with error bars are the SPIRE photometry using
the appropriate fractional contributions from the identified major optical
counterparts, and the blue curves are the best models from CE01. The derived
$L_{IR}$ values are labeled. We also calculate the peak dust temperatures 
($T_{peak}$) based on the best-fit models, which are also labeled.
}
\label{fig:fig_FIR_fit}
\end{figure*}

\begin{deluxetable*}{cccccccccc}
\tablecaption{Properties of the 22 Strongly Variable IDF IRAC Sources Studied in This Work}
    \tablehead{\colhead{IAU Name} & \colhead{ID} & \colhead{Sample} & \colhead{$z_{ph}$} & \colhead{$z_{ph}$-Source} & \colhead{Log~$L_{IR}$} & \colhead{Log~$L_{X}$} & \colhead{$i^{\prime}$} & \colhead{Ch1} & \colhead{Ch2} \\
        &              &                  &                    &                           &   $(\mathrm{L_{\odot}})$ & $(\mathrm{erg/s})$ & & & }
\startdata
IDFV J174002.2+685305.53 & M9038 & X-ray & 0.60 & SDSS & \nodata & 43.5 & 21.02 & 19.13 & 19.21 \\
IDFV J174027.4+685317.25 & M9464 & X-ray & 2.10 & LP-Q & \nodata & 44.6 & 20.92 & 19.16 & 18.89 \\
IDFV J173959.0+685343.79 & M10516 & X-ray & 0.64 & EAZY & \nodata & 43.4 & 21.55 & 20.05 & 19.72 \\
IDFV J174027.5+685450.07 & M13152 & X-ray & 0.82 & SDSS & \nodata & 43.4 & 21.71 & 19.36 & 19.16 \\
IDFV J173922.5+685531.91 & M14706 & FIR/X-ray & 2.05 & LP-Q & 12.8 & 44.3 & 20.99 & 19.90 & 19.52 \\
IDFV J173955.0+685617.78 & M16397 & X-ray & 0.13 & EAZY & \nodata & 42.0 & 20.60 & 19.52 & 19.19 \\
IDFV J174048.5+685702.79 & M18291 & X-ray & 0.69 & EAZY & \nodata & 43.2 & 21.17 & 20.48 & 20.04 \\
IDFV J174021.0+685817.00 & M21355 & FIR & 0.48 & SDSS & 11.4 & \nodata & 20.12 & 19.49 & 19.63 \\
IDFV J173936.1+690013.98 & M25810 & X-ray & 0.73 & EAZY & \nodata & 42.9 & 22.32 & 20.75 & 20.07 \\
IDFV J173955.6+690020.63 & M26166 & FIR/X-ray & 0.29 & SDSS & 11.4 & 42.6 & 18.28 & 18.09 & 18.22 \\
IDFV J174100.6+690027.72 & M26428 & X-ray & 2.30 & LP-Q & \nodata & 44.7 & 23.54 & 21.27 & 21.05 \\
IDFV J174003.3+690048.26 & M27189 & X-ray & 0.74 & SDSS & \nodata & 43.2 & 22.62 & 20.98 & 20.66 \\
IDFV J174031.3+690158.70 & M30165 & X-ray & \nodata & \nodata & \nodata & \nodata & 23.17 & 18.96 & 18.67 \\
IDFV J173930.4+690159.65 & M30195 & X-ray & \nodata & \nodata & \nodata & \nodata & 22.07 & 19.53 & 19.44 \\
IDFV J174012.9+690223.89 & M31099 & X-ray & \nodata & \nodata & \nodata & \nodata & 22.56 & 20.32 & 20.01 \\
IDFV J173901.1+690234.07 & M31465 & X-ray & \nodata & \nodata & \nodata & \nodata & 22.21 & 19.72 & 19.85 \\
IDFV J174116.7+690241.61 & M31844 & X-ray & 0.35 & SDSS & \nodata & 43.7 & 18.96 & 17.94 & 17.73 \\
IDFV J173951.9+690302.22 & M32603 & X-ray & \nodata & \nodata & \nodata & \nodata & 22.69 & 19.70 & 19.51 \\
IDFV J173924.3+690318.01 & M33366 & X-ray & \nodata & \nodata & \nodata & \nodata & 22.89 & 20.72 & 20.61 \\
IDFV J174031.1+690332.22 & M33768 & FIR/X-ray & \nodata & \nodata & \nodata & \nodata & 21.15 & 18.48 & 18.47 \\
IDFV J174026.8+690353.12 & M34610 & X-ray & 0.40 & LP-Q & \nodata & 42.1 & 21.04 & 19.37 & 19.15 \\
IDFV J173855.7+690451.92 & M36871 & X-ray & 1.00 & LP-Q & \nodata & 44.4 & 20.41 & 18.72 & 18.50
\enddata
\tablecomments{Properties of the 22 strongly variable IDF IRAC sources studied
in this work. The columns are: (1) IAU name, where the coordinates are based
on the OID $i^{\prime}$ positions, (2) internal ID used in our IDF catalog,
(3) sample(s) to which a source belongs, i.e., FIR sample, X-ray sample, or
both, (4) $z_{ph}$, (5) the source of $z_{ph}$, i.e., from the SDSS, our own
EAZY run using galaxy templates or LePhare run using QSO/AGN templates (LP-Q),
(6) $log(L_{IR})$, (7) $log(L_X)$, (8) $i^{\prime}$ magnitude as measured
in the ODI image, (8) averaged Ch1 magnitude after aperture correction, and (9)
averaged Ch2 magnitude after aperture correction.
}
\end{deluxetable*}

\subsection{IR and X-ray Luminosities}

   We then determine whether these 22 variable objects reside in (U)LIRGs
and/or AGNs, for which we need their redshifts to calculate their IR and/or
X-ray luminosities. None of them have spectroscopic redshifts, and we adopt
six photometric redshifts ($z_{ph}$) available from the SDSS DR14. For the rest,
we derive their $z_{ph}$ on our own by fitting their spectral energy 
distributions (SEDs). To best match the existing SDSS $z_{ph}$, we confine
these SEDs to the optical regime and use the SDSS photometry. Seven objects
have to be excluded from this analysis because they are not detected in the SDSS
due to their faintness, one of which is in both the FIR and the X-ray samples
and the other six are in the X-ray sample. In the end, we have nine objects
that need their $z_{ph}$ derived. 

   An X-ray AGN does not necessarily show AGN signatures in other wavelengths.
Therefore, we first treat these nine objects as ``normal'' galaxies dominated
by starlight in the optical, and fit them using the EAZY software 
\citep[][]{Brammer2008}. An important reason to use EAZY is that it allows
Bayesian prior(s) to assign very low weight to unreasonable $z_{ph}$ solutions,
which is particularly important in our case because galaxies bright enough to be
detected in the SDSS should be at $z\lesssim 0.8$. Specifically, we use its
\texttt{prior\_R\_extend.dat}, and treat the SDSS $r^{\prime}$-band as the
$R$-band
\footnote{Such a prior sets the likelihood of redshift for a given magnitude;
as an example, the adopted function at $m=20$~mag peaks at $z=0.24$.}.
We obtain a satisfactory fit for four of them. 
For the five objects that do not fit well from the above, we consider the 
possibility that their optical light is dominated by AGN. We fit them using
the LePhare software \citep[][]{Arnouts1999,Ilbert2006}, which includes QSO/AGN 
templates. We note that no prior is applied in the LePhare run. As it turns 
out, all these five objects can be fit reasonably well.

   To check the consistency of our $z_{ph}$ with the SDSS results, we also
repeat the same procedure for the aforementioned six objects that already have
SDSS $z_{ph}$. We find that they can all be well fit using EAZY with galaxy 
templates, and the differences between the SDSS $z_{ph}$ and ours have mean 
$<$$\Delta z/(1+z)$$>$$=0.04$. Therefore, we believe that combining these two
sets of $z_{ph}$ is reasonable. Figure 4 summarizes the SED fitting results for
all these 15 objects.

   For those in the FIR sample, we construct their FIR SEDs using the SPIRE
photometry based on the fractional contributions of the major contributors, and
derive $L_{IR}$ by fitting to the templates of \citet[][]{CE2001}, following
\citet[][]{MY15}. M33768 is not detected in the SDSS and hence is not
considered. Figure 5 shows the FIR SED fitting results for the other three 
objects. All four IRAC variable objects in the FIR sample have 
$L_{IR}>10^{11}L_\odot$, qualifying as (U)LIRGs.

  For those in the X-ray sample, we calculate their X-ray luminosities
($L_{X}$) based on the \Chandra\, photometry. This is done using
$L_{X}=f_X\times 4\pi D_L^2$, where $f_X$ is the flux density over rest-frame
0.2--10~keV and $D_L$ is the luminosity distance.  To obtain $f_X$, a power-law
SED in the form of $I_\nu \approx \nu^{-\alpha}$ is fit to the flux densities
at different energy bands, and the best fit is integrated over rest-frame
0.2--10~keV. We find that they all have 
$L_{X}\geq 10^{42}$~erg s$^{-1}$ cm$^{-2}$ 
and thus fall within the nominal range of X-ray AGNs. 

  Table 1 summarizes the properties of all these objects discussed above.
For the sake of completeness, the table also lists those that do not yet have
$z_{ph}$ estimates. These latter objects are also included in the discussions
below as potential (U)LIRGs and AGNs.

\section{Interpretation}

  While it is natural to attribute the variabilities seen in the X-ray sample
to AGN variability, our focus here is to examine whether the four variable 
objects in the FIR sample (see Figure 2) could be due to multiple SNe. 

\subsection{SNe versus AGNs}

  It is plausible that the variability of M21355 is caused by multiple SNe, as
it is the only non-X-ray source (among the four) and hence very
likely does not harbor an AGN. However, the other three need
more examination because they are X-ray AGNs as well.

  We suggest that the variability of M26166 is also likely due to multiple SNe
for two reasons. First, both M26166 and M21355 have blue
Ch1$-$Ch2 colors. This means that their 3--5 $\mu$m emission cannot be dominated
by AGNs \citep[e.g.,][]{Stern2005}, which then leaves multiple SNe
as the plausible explanation for their variabilities. In contrast, most other
variables in the X-ray sample (15 out of 18) have red Ch1$-$Ch2 colors (i.e.,
consistent with AGN SEDs), and only three are exceptions (M9038, M31466, and
M33366). 

  Second, both M26166 and M21355 have a long ``quiet'' 
phase ($\gtrsim 6$ years), which is different from those of the other 19 
variable objects in the two samples combined. We argue that this is actually an
expected feature in our current search. If a (U)LIRG maintains a constant, high 
$r_{SN}$ all the time, the events cannot be easily detected from the
variability because the host is always at an elevated flux
level due to the SNe overlapping in time.
On the other hand, a (U)LIRG could also achieve the same
average rate by erupting a few times more SNe over a short period 
and then remaining ``quiet'' over a period longer by the same factor. In
this case, the variability becomes more significant,
which could be what we see in M21355 and M26166.

   By these arguments, the variabilities of M14706 and M33768 in the FIR sample
are most likely due to AGNs, like others in the X-ray sample. In particular,
this should not be surprising for M14706, as it is one of the five objects 
whose optical SEDs are better fit with quasar/AGN templates, and should be a 
quasar based on its point-like morphology (see Figure 2).


\subsection{Consistency with SNe Cause}

   Accepting that M21355 and M26166 could have multiple SNe, we now examine how
well their light curves can be explained. Unfortunately, this cannot be done
quantitatively with the current data, because we do not know the types of SNe
involved, let alone their relative fractions. Nevertheless, we can still
qualitatively examine whether the amplitudes and the durations of the 
variabilities could be consistent with the SNe interpretation.

   Figure 6 shows their ``net'' Ch1 variations after subtracting the host fluxes
as determined over the ``quiet'' periods. 
We model the variations by composing
the average $K$-band light curve template of Type II SNe derived by 
\citet[][]{Mattila2001}, which is represented by two power-law functions before
and after the maximum. While this approach does not capture the detailed
behaviors of the real light curves, the template is sufficient for such a
toy model. The average peak absolute magnitude of this template
is $K=-18.6$~mag in Vega system, or equivalently, $K_{AB}=-16.73$~mag.

   At $z\approx 0.3$--0.5, Ch1 is close to the rest-frame $K$-band; while it
samples slightly longer wavelengths than $K$-band, we ignore this small
difference.  The modeling is done by combining an arbitrary number of 
light curve templates (time-dilated at the source redshift) at arbitrary times.
This results in a large number of plausible model sets, one of which is shown
in Figure 6 for each source to explain one of the most prominent features. The 
features span over $\sim$800 days in the observer's frame, and can be reasonably
modeled by a stack of 60 and 80 templates for M21355 and M26166, respectively.
This translates to $r_{SN}$ of $\sim$40--47 events~yr$^{-1}$ (in rest-frame),
which is qualitatively consistent with the expectation that we are observing
an elevated rate over a short period of time as discussed in \S 4.1 above. 

\begin{figure*}[t]
\plotone{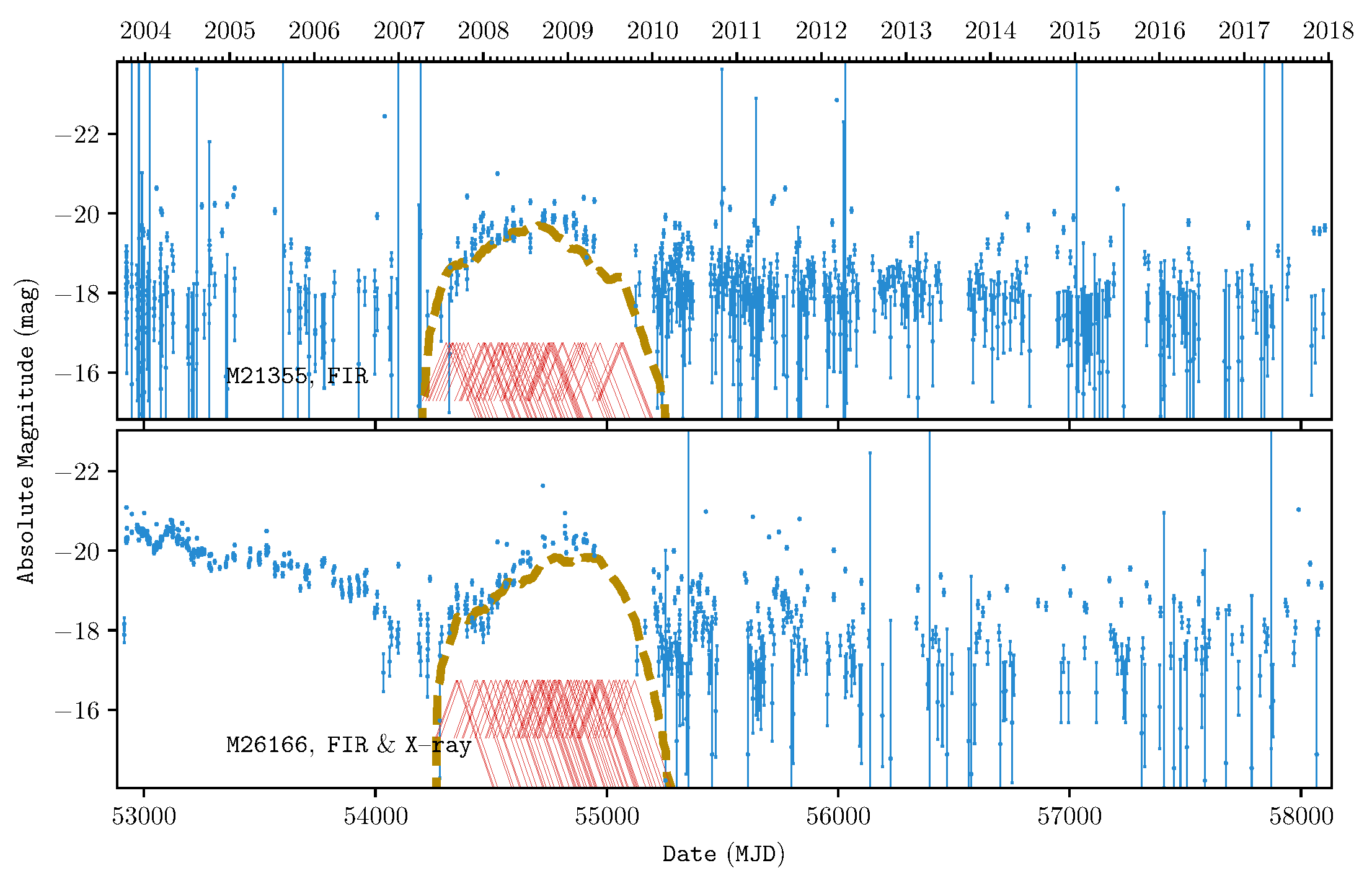}
\caption{Net Ch1 variabilities of M21355 and M26166 derived by subtracting the
host light based on their quiet phases (after 2011 for both; see Figure 2) and 
toy models that qualitatively explain the features of the light curves. The
blue dots with error bars (calculated based on the $S/N$ after subtraction of
the host) are the data points, while the horizontal dashed gray line indicates
the ``zero-flux'' level around which there are equal number of positive and
negative residuals in the quiet phases (the negative residuals cannot be shown
due to the magnitude scale). 
For clarity, the toy models are only shown for
one of the most prominent features for each source. The models are constructed
using the power-law $K$-band light curve template for Type II SNe as derived by
\citet[][]{Mattila2001}. The features, which span over $\sim$800 days in the
observer's frame, can be qualitatively explained by the combined effects
(dashed golden curves) resulted from stacking 60 and 80 templates (red curves)
for M21355 and M26166, respectively.
}
\label{fig:netlc}
\end{figure*}

   We emphasize that such a toy model is only meant to demonstrate that the
features in the light curves can be explained by multiple SNe but not to
deduce any detailed properties of the SNe. 
Many important factors have to be omitted because the current data do not
warrant the consideration of such details. For example, we do not consider the
population of ``super luminous SNe'' (at least two magnitudes more luminous
than the adopted template) that have been known for two decades 
\citep[][]{GalYam2012}, which would decrease the required $r_{SN}$. We also
neglect dust extinction in near-IR, which could still be significant and could
affect the brightness of the involved SNe.
Nevertheless, it is particularly encouraging that
using just one specific SN light curve can explain the features, and one
can imagine that allowing different types of SNe would only work better.

\section{Discussion}

   Strictly speaking, the analysis in \S 4 only shows that the variabilities in
M21355 and M26166 are consistent with the SNe interpretation but does not 
definitely prove it, the latter of which probably could only be claimed if
the SNe were detected individually. However, we are able to rule out the 
AGN variability as their cause based on the Ch1/2 color of the hosts and the
light curve behaviors, and hence leave the SNe explanation as the likely
alternative. Multiple SNe have been detected individually in at least two local
LIRGs (Arp 299 and IC 833; see the references in \S 1), and thus it should not
be surprising that we find evidence of similar events at high redshifts.

   One might question why such variabilities cannot be due to other types of
transients. Among all other populations, only tidal disruption events
\citep[TDEs; see][,for a review]{Komossa2015}, which are thought to be due
to the disruption of a star falling into the supermassive black hole at a
galaxy center, could possibly have IR amplitude and varying time scale
comparable to SNe. TDEs emit most strongly in the X-ray to optical wavelengths,
and these energetic photons could be absorbed by the dusty ISM around the black
hole and re-emit in IR \citep[see, e.g.][]{Lu2016}. Such IR ``echoes'' have
been found for a few TDEs \citep[][]{Dou2016, Dou2017, Jiang2016, Jiang2017, 
vanVelzen2016} by using the {\it Wide-field Infrared Survey Explorer}
({\it WISE}) data \citep[][]{Wright2010} in $W1$ (3.4~\micron) and $W2$ 
(4.6~\micron) bands, which are close to the Ch1/2
bands used in our study. \citet[][]{Wang2018} further present a sample of 
{\it WISE}
sources that have similar variability and suggest that they could also be the
IR echoes of TDEs (but see also \citet[][]{Assef2018}). 
More energetic TDEs have also been suggested, such as the very luminous
transient in Arp 299 recently discovered by \citet[][]{Mattila2018}.
However, invoking TDEs
for M21355 and M26166 is problematic, because both objects would certainly
require multiple TDEs to explain the features in their light curves. This would
imply several events over ten years, which is orders-of-magnitude higher than
the expected rate of $\sim$10$^{-5}$~yr$^{-1}$ per galaxy 
\citep[][]{Wang2004, Wang2012, vanVelzen2014, Holoien2016}. 
\footnote{While there are now suggestions that TDEs could occur at a much
higher rate in (U)LIRGs \citep[][]{Tadhunter2017}, such a connection is yet to
be established.}
Therefore, we believe that TDEs are unlikely the cause. In contrast, the 
multiple SNe interpretation comes naturally, as it meets the expectation that
(U)LIRGs should have a high rate of SNe because of their high SFR. 

   We emphasize that our pilot study here is the first one using variability of
integrated IR light to reveal dust-embedded SNe in (U)LIRGs beyond the local
universe. There have been other IR variability studies in the literature, some
of which also utilize IRAC Ch1/2. A significant example is that of
\citet[][]{Kozlowski2010b}, who investigated the IRAC variability of objects in
the 8.1~deg$^2$ \Spitzer\, Deep Wide-Field Survey in the Bo\"{o}tes field
over four epochs spanning four years. Aiming at variability in general,
however, they do not target (U)LIRGs, and most of their variable objects are
due to AGNs. While this work has led to the serendipitous discovery of a 
self-obscured SN at $z\approx 0.2$ \citep[][]{Kozlowski2010c}, the SN is not
related to a (U)LIRG and its obscuration is due to its dusty circumstellar
medium but not the environment. Another example is the ongoing SPitzer InfraRed 
Intensive Transients Survey (SPIRITS; \citet[][]{Kasliwal2017}), which searches
for Ch1/2 transients in 190 nearby galaxies. SPIRITS has also
led to the discovery of new SNe: \citet[][]{Jencson2017} report two SNe in IC
2163 (not a LIRG) separated by less than two years. This survey, which does not
include (U)LIRGS, is confined to the local universe ($< 15$~Mpc) and only aims
to discover transients that can be individually resolved.

   The implication of our study can be understood in two-fold.
First, we present strong supporting evidence that high-z (U)LIRGs, like their
local counterparts, have high rate of SNe. While determining $r_{SN}$ is
beyond the scope of this paper, it is obvious that the two cases
\footnote{We note that the other objects in the FIR sample also have hints of
variabilities but are at lower levels; however we have to differ the discussion
to a forthcoming paper because our current analysis does not yet allow us to
confidently assess these more subtle features in the light curves.}
shown in Figure 4 require multiple SNe per year over the active periods. This
makes (U)LIRGs ideal targets to search for high-z SNe in the rest-frame IR,
which will become feasible when the \JWST\, comes online in the near future.
The unprecedented IR resolution and sensitivity offered by the \JWST\, will
easily enable us to detect such SNe individually out to any redshifts where
(U)LIRGs are seen and to assemble large samples that can lead to many
applications. Second, using the integrated light variability as the indicator
of SNe in high-z (U)LIRGs will remain relevant
in the \JWST\, era because it will still be difficult for \JWST\, to
probe close to the nuclear regions. For example, \JWST's resolution at
4.4~\micron\, (the reddest band of its NIRCam instrument) is 
$\sim $ 0\arcsec.17, which corresponds to $\sim 1.4$~kpc at
$z\approx 1$--2. ULIRGs have now been found out to $z\approx 6$--7
\citep[][]{Riechers2013, Fudamoto2017, Strandet2017}, and to sample the
rest-frame $K$-band at such redshifts \JWST\, will need to observe at 
18--20\micron\, using its MIRI instrument at the resolution of
$\sim $ 0\arcsec.75, which corresponds to $\sim 4.2$~kpc.
In such cases, our method will be the only option. Of course, one would not
be able to obtain \JWST\, time baseline or cadence comparable to the IDF IRAC
observations, and the ``quiet phase'' argument presented in \S 4 would not be 
applicable. However, if the \JWST\, monitoring is done in at least two bands,
the color information can always be used to judge if the observed variability
is more likely due to SNe or AGN. In addition, contemporary observations
in one optical band can also greatly help the judgment, because the 
variability caused by dusty SNe should not be seen in optical due to the
large extinction in (U)LIRGs.

   Lastly, we point out that the very field of IDF is of great interest for
the \JWST. By design, the IDF is close to the North Ecliptic Pole and
thus is in the continuous viewing zone (CVZ) of \JWST, which is the narrow
region within $\pm 5^{\circ}$ from the Ecliptic Poles
\footnote{See ``James Webb Space Telescope User Documentation'',
\url{https://jwst-docs.stsci.edu/display/JTI/JWST+Observatory+Coordinate+System+and+Field+of+Regard}
}.
For this reason, the IDF can be visited by the \JWST\, at any time of the year.
It will be ideal for a \JWST\, monitoring program, especially
when considering the fact it is the only region in the CVZ that has
deep \Herschel\, data revealing a large sample of high-z (U)LIRGs.

\section{Summary}


   If (U)LIRGs are mainly driven by starbursts, they should have high rate
of SNe embedded by dust. This would make them ideal ``SNe factories'' that can
produce large SNe samples at high-z if the search is done in the rest-frame IR
where the dust extinction is minimal. In this study, we use the IDF 3--5~$\mu$m
data, which span more than 14 years, to test this idea. We propose that such 
dust-embedded SNe in (U)LIRGs can be revealed by the variabilities in the 
integrated near-IR light of the host galaxy, which can be applied to high-z
where it is difficult to discern SNe individually due to the lack of sufficient
spatial resolution and/or sensitivity. Our paper demonstrates the feasibility
of this method. Out of the 96 potential high-z (U)LIRGs that have the best
temporal coverage (the FIR sample), we identify four strongly variable objects.
We show that the contamination due to AGN variability can be 
discriminated based on information such as their colors, and present
two strong cases that are consistent with multiple SNe overlapping in time.
It is very likely that a future \JWST\, monitoring program targeting fields
of known (U)LIRGs can resolve many such SNe. However, variability study will
remain the only way to probe the SN activities close to the nuclear regions
where even \JWST\, still lacks the resolution to resolve the SNe individually.

\acknowledgments{
  We acknowledge the support of the University of Missouri Research Board Grant
RB 15-22 and NASA's ADAP Program under grant number NNX15AM92G. JFB is 
supported by NSF Grant No. PHY-1714479. Part of our data processing and 
analysis were done using the HPC resources at the University of Missouri 
Bioinformatics Consortium (UMBC) and the Ohio Supercomputer Center (OSC).
}


\end{document}